\documentclass[english,aps,preprint,superscriptaddress]{revtex4-2}
\usepackage{lmodern}
\usepackage{lmodern}
\usepackage[T1]{fontenc}
\usepackage[utf8]{inputenc}
\usepackage{amsmath}
\usepackage{amsthm}
\usepackage{amssymb}
\usepackage{graphicx}
\usepackage{esint}

\makeatletter

\usepackage{amsthm}\usepackage{latexsym}\usepackage{bm}\usepackage{amsfonts}

\usepackage{babel}
\usepackage{hyperref}
\usepackage{enumitem}

\usepackage{hyperref}
\hypersetup{
	breaklinks = true,
    colorlinks = true,
    citecolor = {blue},
	urlcolor = {blue},
	linkcolor = {blue}
}

\makeatother

\usepackage{babel}
\begin{document}
\title{Spontaneous and explicit parity-time-symmetry breaking in drift wave
instabilities}
\author{Hong Qin}
\email{hongqin@princeton.edu}

\affiliation{Plasma Physics Laboratory, Princeton University, Princeton, NJ 08543,
U.S.A}
\author{Yichen Fu}
\affiliation{Plasma Physics Laboratory, Princeton University, Princeton, NJ 08543,
U.S.A}
\author{Alexander S. Glasser}
\affiliation{Plasma Physics Laboratory, Princeton University, Princeton, NJ 08543,
U.S.A}
\author{Asher Yahalom}
\affiliation{Ariel University, Kiryat Hamada POB 3, Ariel 40700, Israel}
\affiliation{Plasma Physics Laboratory, Princeton University, Princeton, NJ 08543,
U.S.A}
\begin{abstract}
A method of Parity-Time (PT)-symmetry analysis is introduced to study
the high dimensional, complicated parameter space of drift wave instabilities.
We show that spontaneous PT-symmetry breaking leads to the Ion Temperature
Gradient (ITG) instability of drift waves, and the collisional instability
is the result of explicit PT-symmetry breaking. A new unstable drift
wave induced by finite collisionality is identified. It is also found
that gradients of ion temperature and density can destabilize the
ion cyclotron waves when PT symmetry is explicitly broken by a finite
collisionality. 
\end{abstract}
\maketitle
Drift wave instability is an important research topic for plasma physics
and magnetic fusion \citep{Rudakov60,Krall1963,Tang1978,horton1999drift,Horton2017}.
The turbulence transport in tokamaks starts from the seed of drift
waves destabilized by various effects that naturally exist in the
devices. The drift waves in tokamaks typically depend on 7 or 8 dimensionless
parameters, which results in many different paths to instabilities,
including the Ion Temperature Gradient (ITG) mode \citep{Coppi1967,parker1993gyrokinetic},
the universal mode \citep{Krall1965}, the electron temperature gradient
mode \citep{Dorland00-5579,Dimits2007}, and the collisional mode
\citep{Hendel1968}, just to name a few. The parameter dependency
of drift wave instabilities is complex. 

In the present study, we introduce a new perspective to understand
the mechanism of drift wave instabilities using the tool of Parity-Time
(PT)-symmetry analysis. PT symmetry is a concept in non-Hermitian
quantum physics \citep{Dirac1942,Pauli1943,Lee1969} introduced by
Bender and collaborators \citep{bender1998real,bender2002complex,bender2007making,Bender2010,mostafazadeh2002pseudo,mostafazadeh2002pseudoII,mostafazadeh2002pseudoIII,Zhang2020PT}.
It has found a wide range of applications in many branches of physics.
It was recently applied to study instabilities in continuous media,
such as the classical Kelvin-Helmholtz instability and the Rayleigh-Taylor
instability \citep{Qin2019KH,Fu2020KH}. Here, using a two-fluid model
in a slab geometry, we show that drift wave instabilities in magnetized
plasmas can be divided into two classes that are induced by spontaneous
PT-symmetry breaking and explicit PT-symmetry breaking, respectively. 

Spontaneous PT-symmetry breaking happens in conservative systems via
Krein resonances \citep{Zhang2020PT,Krein50,Gelfand55,Yakubovich75}
between two eigenmodes with opposite signs of actions. Explicit PT-symmetry
breaking occurs in non-conservative systems, and it is often associated
with dissipative instabilities \citep{Kirillov2013a}. Because the
spectrum of a PT symmetric system must be symmetric with respect to
the real axis, instabilities arise when and only when an eigenmode
violates  PT symmetry, i.e., PT symmetry is spontaneously broken.
As a consequence, spontaneous PT-symmetry breaking usually has a finite
threshold in the parameter space corresponding to the Krein resonance.
If a physical effect is introduced into the system such that it does
not admit PT symmetry anymore, we say that PT symmetry is explicitly
broken. For such a system, the constraints on the distribution of
the spectrum are removed, and instabilities are easier to trigger.
When explicit PT-symmetry breaking destabilizes the system, in most
cases, there is no finite threshold for the onset of instabilities. 

We will use the example of the ITG instability and the collisional
instability to demonstrate the physics of spontaneous and explicit
PT-symmetry breaking in drift waves, respectively. In particular,
we show that the governing equation of the ITG mode is PT symmetric,
and spontaneous PT-symmetry breaking leads to the ITG instability.
The finite threshold of the ITG mode corresponds to the Krein resonance
for spontaneous PT-symmetry breaking. On the other hand, the collisions
between electrons and ions explicitly breaks PT symmetry, and there
is no threshold for the collisional instability. When the gradient
of temperature or density exists, any small collisionality will induce
a growth rate for the drift waves. A finite collisionality also induces
a new unstable low-frequency drift wave, which will be temporarily
called collision-induced drift wave for lacking of a better terminology. 

In addition, within the model adopted, we found that temperature gradient
and density gradient are destabilizing for the electrostatic cyclotron
waves as well. Because of the strong constraint of PT-symmetry, the
threshold for instability is too high for the parameters of practical
interest. However, when PT symmetry is explicitly broken by a small
but finite collisionality, the gradients of ion temperature and density
can derive the electrostatic cyclotron waves unstable without any
threshold.

We first demonstrate the mechanism of spontaneous PT-symmetry breaking
for the local ITG mode in a slab geometry using a conservative two-fluid
model. For each species, the governing equations are 
\begin{align}
\frac{\partial n_{j}}{\partial t}+\nabla\cdot\left(n_{j}\boldsymbol{u}_{j}\right) & =0\thinspace,\label{eq:continuity}\\
m_{j}n_{j}\left(\frac{\partial\boldsymbol{u}_{j}}{\partial t}+\boldsymbol{u}\cdot\nabla\boldsymbol{u}\right) & =Z_{j}n_{j}e\left(\boldsymbol{E}+\frac{\boldsymbol{v}\times\boldsymbol{B}}{c}\right)-\nabla p_{j}\thinspace,\label{eq:momentum}\\
\frac{d}{dt}\left(\frac{p_{j}}{\left(m_{j}n_{j}\right)^{\gamma_{j}}}\right) & =0\thinspace,\label{eq:energy}
\end{align}
where $Z_{j}$ is charge number and $\gamma_{j}$ is the polytropic
index for the $j$-th species. 

The equilibrium consists of a constant magnetic field in the $z$-direction,
$\boldsymbol{B}=B\boldsymbol{e}_{z}$, and inhomogeneous density $n_{j0}(x)$
and pressure $p_{j0}(x)=n_{j0}(x)T_{j0}(x)$ of electrons and ions.
The gradients of densities and pressures are in the $x$-direction.
Quasi-neutrality condition requires $Z_{i}n_{i0}=n_{e0}$. Because
the equilibrium is inhomogeneous, particles flow in the diamagnetic
direction. We assume that an equilibrium electrostatic field in the
$x$-direction is established such that only electrons flow in the
diamagnetic direction, i.e., 
\begin{align}
\boldsymbol{u}_{i0} & =0\thinspace,\label{eq:ui0}\\
\boldsymbol{u}_{e0} & =-\left(\frac{cE_{0}(x)}{B}+\frac{c}{eBn_{e0}}\frac{dp_{e0}(x)}{dx}\right)\boldsymbol{e}_{y}\thinspace,\label{eq:ue0}\\
\boldsymbol{E}_{0}(x) & =E_{0}(x)\boldsymbol{e}_{x}=\frac{1}{Z_{i}n_{i0}(x)e}\frac{dp_{i0}(x)}{dx}\boldsymbol{e}_{x}\thinspace.\label{eq:E0}
\end{align}

For linear electrostatic perturbation, we consider a local mode in
the form of $\exp(ik_{y}y+ik_{z}z-i\omega t).$ For the ITG mode,
electron density perturbation can be approximated by the adiabatic
response (see Appendix \ref{sec:Non-adiabatic-response-of}), 
\begin{equation}
Z_{i}n_{i1}(x)=n_{e1}(x)=\frac{en_{e0}\phi_{1}}{m_{e}T_{e0}}\,,\label{eq:QN}
\end{equation}
where the first equal sign is due to the quasi-neutrality condition.
Ions' response is governed by the linearized system of Eqs.\,(\ref{eq:continuity})-(\ref{eq:energy}), 

\begin{align}
-i\omega n_{i1} & =-ik_{z}n_{i0}u_{iz1}-ik_{y}n_{i0}u_{iy1}-u_{ix1}\frac{dn_{i0}}{dx}\,,\label{ni1}\\
-i\omega u_{ix1} & =\frac{Z_{i}eu_{iy1}B_{0}}{cm_{i}}+\frac{1}{m_{i}n_{i0}^{2}}\frac{dp_{0}}{dx}n_{i1}\thinspace,\\
-i\omega u_{iy1} & =-\frac{ik_{y}Z_{i}e\phi_{1}}{m_{i}}-\frac{Z_{i}eu_{ix1}B_{0}}{cm_{i}}-\frac{ik_{y}p_{i1}}{m_{i}n_{i0}}\thinspace,\\
-i\omega u_{iz1} & =-\frac{ik_{z}Z_{i}e\phi_{1}}{m_{i}}-\frac{ik_{z}p_{i1}}{m_{i}n_{i0}}\thinspace,\label{uiz1}\\
-i\omega p_{i1} & =-u_{ix1}\frac{dp_{i0}}{dx}-\gamma_{i}p_{i0}(ik_{y}u_{iy1}+ik_{z}u_{iz1})\thinspace.\label{pi1}
\end{align}
We choose the following normalization and dimensionless parameters,
\begin{flalign}
\bar{t} & =t\Omega_{i},\,\bar{x}=\frac{x}{a},\thinspace\bar{k}_{y,z}=k_{y,z}a,\thinspace\bar{k}_{n}=\frac{1}{n_{i0}}\frac{dn_{i0}}{dx}a\,\thinspace\thinspace,\label{nor1}\\
\bar{u}_{ix1,iy1,iz1} & =\frac{u_{ix1,iy1,iz1}}{a\Omega_{i}}\thinspace,\bar{v}_{thi}^{2}\equiv\frac{T_{i0}}{m_{i}a^{2}\Omega_{i}^{2}},\thinspace\bar{v}_{s}^{2}\equiv\frac{Z_{i}T_{e0}}{m_{i}a^{2}\Omega_{i}^{2}},\thinspace\bar{p}_{i1}=\frac{p_{i1}}{n_{i0}m_{i}\Omega_{i}^{2}a^{2}}\,,\label{nor2}\\
\bar{\omega}_{T}^{\dagger} & \equiv\frac{1}{m_{i}a\Omega_{i}^{2}}\frac{dT_{i0}}{dx},\thinspace\bar{\omega}_{p}^{\dagger}\equiv\frac{1}{m_{i}n_{i0}a\Omega_{i}^{2}}\frac{dp_{i0}}{dx}=\bar{k}_{n}\bar{v}_{thi}^{2}+\thinspace\bar{\omega}_{T}^{\dagger}\,.\label{nor3}
\end{flalign}
Here, $\Omega_{i}=Z_{i}eB/m_{i}c$ is the ion gyro-frequency, and
$a$ is the typical scale length of the system, which can be chosen
to be the minor radius of a tokamak. The dimensionless parameter $\bar{k}_{n}$
measures the density gradient, and $\bar{\omega}_{T}^{\dagger}$ measures
the ion temperature gradient. Substituting Eq.\,(\ref{eq:QN}), we
cast the linear system Eqs.\,(\ref{ni1})-(\ref{pi1}) into the form
of Schrödinger's equation,
\begin{align}
H\psi & =\omega\psi\thinspace,\label{Hpsi}\\
H & =\left(\begin{array}{ccccc}
0 & -ik_{n} & k_{y} & k_{z} & 0\\
i\omega_{p}^{\dagger} & 0 & i & 0 & 0\\
v_{s}^{2}k_{y} & -i & 0 & 0 & k_{y}\\
v_{s}^{2}k_{z} & 0 & 0 & 0 & k_{z}\\
0 & i\omega_{p}^{\dagger} & \gamma_{i}v_{thi}^{2}k_{y} & \gamma_{i}v_{thi}^{2}k_{z} & 0
\end{array}\right)\thinspace,\label{H}\\
\psi & =\left(n_{i1},u_{ix1},u_{iy1},u_{iz1},p_{i1}\right)^{T}\thinspace.\label{psi}
\end{align}
All quantities in Eqs.\,(\ref{Hpsi})-(\ref{psi}) are normalized
and dimensionless. For easy notation, the over bars for normalized
quantities have been dropped. This convention will be adopted hereafter
unless explicitly stated otherwise. 

The spectrum of the system is determined by the characteristic polynomial
of $H$,
\begin{align}
D\left(\omega\right)= & -\omega^{5}+\alpha\omega^{3}+\beta\omega^{2}+\xi\omega=0\thinspace,\label{DR}\\
\alpha\equiv & 1+k_{n}^{2}v_{thi}^{2}+\left(k_{y}^{2}+k_{z}^{2}\right)\left(v_{s}^{2}+\gamma_{i}v_{thi}^{2}\right)+k_{n}\omega_{T}^{\dagger}\thinspace,\\
\beta\equiv & k_{y}\left(2\omega_{T}^{\dagger}+k_{n}\left(2v_{thi}^{2}+v_{s}^{2}\right)\right)\thinspace,\\
\xi\equiv & -k_{n}^{2}\left(k_{y}^{2}+k_{z}^{2}\right)v_{thi}^{4}(\gamma_{i}-1)-k_{z}^{2}\left(v_{s}^{2}+\gamma_{i}v_{thi}^{2}\right)\thinspace\\
 & -k_{n}\left(k_{y}^{2}+k_{z}^{2}\right)v_{thi}^{2}(\gamma_{i}-2)\omega_{T}^{\dagger}+\left(k_{y}^{2}+k_{z}^{2}\right)\omega_{T}^{\dagger2}\,.
\end{align}
The spectrum are determined by 7 dimensionless parameters, and consists
of two high-frequency ion cyclotron modes and two low-frequency drift
modes. The zero frequency mode is not physically significant unless
other effects, such as the collisions, are considered. We will come
back to this mode later. 

In a homogeneous equilibrium, $k_{n}=0$, $\omega_{T}^{\dagger}=0$,
and we can set $k_{z}=0$ to observe that the dispersion relation
for the two high-frequency waves reduces to (in un-normalized variables)
\begin{equation}
\omega^{2}=\Omega_{i}^{2}+k_{y}^{2}\left(v_{s}^{2}+\gamma_{i}v_{thi}^{2}\right)\thinspace,
\end{equation}
which is the dispersion relation of the electrostatic ion cyclotron
waves in a homogeneous, magnetized plasma. The dispersion relation
for the two drift waves reduce to $\omega=0$, since the equilibrium
is homogeneous. 

We would like to know when the drift waves, and the electrostatic
ion cyclotron waves, will become unstable. Because the parameter space
is 7 dimensional, the boundary between stability and instability in
the parameter space might be complicated. It turns out that PT-symmetry
analysis can help us to understand how the instability is triggered
in the parameter space. 

Even though Eq.\,(\ref{Hpsi}) assumes the form of Schrödinger's
equation, the Hamiltonian specified by $H$ in Eq.\,(\ref{H}) is
not Hermitian as in standard quantum mechanics. Otherwise, $\omega$
would always be real and the system would be stable. As we show now,
$H$ is PT symmetric instead, which allows instabilities.

Note that when the background is homogeneous, $k_{n}=0$ and $\omega_{T}^{\dagger}=0$,
then $H$ is similar to a Hermitian matrix via a simple rescaling
of variables. This is akin to the situation of cold plasma waves in
a homogeneous medium \citep{Parker2020}. It is the background inhomogeneity
that breaks the Hermiticity of the system, and transform it into a
more interesting PT symmetric system \citep{BenderPrivate2019}. 

\begin{figure}[ptb]
\begin{centering}
\includegraphics[width=5in]{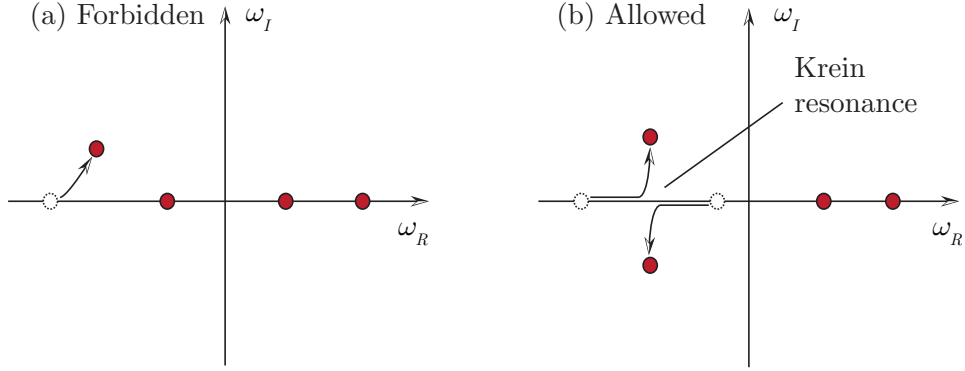}
\par\end{centering}
\caption{Forbidden path (a) and allowed path (b) for instabilities as system
parameters varying for a PT symmetric system, illustrated for systems
with four eigenmodes. A PT symmetric system can be destabilized only
by the spontaneous PT-symmetry breaking through the Krein resonance
(b). }

\label{fig:Krein}
\end{figure}

A non-Hermitian operator $H$ is PT symmetric if $H$ commutes with
an $PT$ operator, i.e.,

\begin{equation}
PTH=HPT\thinspace.\label{eq:PT}
\end{equation}
Here, $P$ is a linear parity operator satisfying $P^{2}=I$ and $T$
is the complex conjugate operation \citep{bender2007making}. We briefly
summarize the properties of PT symmetry as follows. The spectrum of
a PT symmetric operator must be symmetric with respect to the real
axis, which is a strong constraint on the distribution of the eigenmodes.
For a stable system to become unstable under the variation of system
parameters, two eigenmodes of the system must resonate first, which
is called Krein resonance \citep{Krein50,Gelfand55,Yakubovich75,Zhang2020PT}.
An stable eigenmode is forbidden to move away from the real axis by
itself without going through the Krein resonance with another stable
mode. This situation is illustrated in Fig.\,\ref{fig:Krein}.
 It is also known that only Krein resonances between two stable eigenmodes
with opposite signs of actions result in destabilization \citep{Krein50,Gelfand55,Yakubovich75,Zhang16GH,Qin2019LH,Zhang2020PT}.
After resonance between two stable modes with the same sign of actions,
the modes remain stable. Furthermore, when the system is stable, any
eigen vector $\psi$ admits PT symmetry, i.e., $PT\psi=\lambda\psi$
for some complex number $\lambda$. In this case, we say  PT symmetry
is unbroken. When an eigenmode is destabilized after the Krein resonance,
it must also break PT symmetry, i.e., $PT\psi\neq\lambda\psi$ for
any complex number $\lambda$. This is known as spontaneous PT-symmetry
breaking. These eigenmode properties have been identified for applications
in plasma physics and beam physics \citep{Qin14-044001,Qin15-056702,Chung15,Zhang16GH,Fraser2018,Qin2019KH,Fu2020KH}. 

\begin{figure}[ptb]
\begin{centering}
\includegraphics[width=3in]{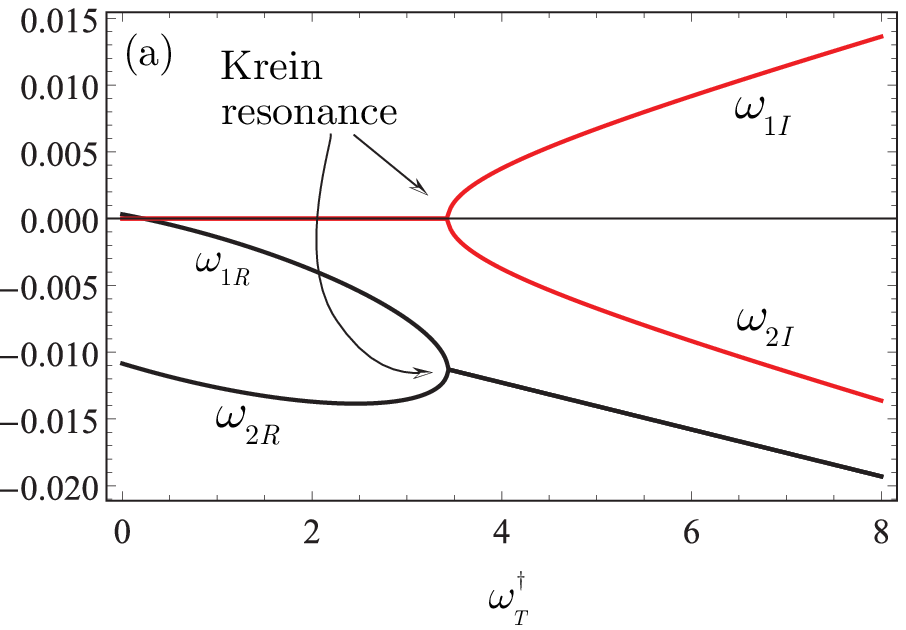}\includegraphics[width=3in]{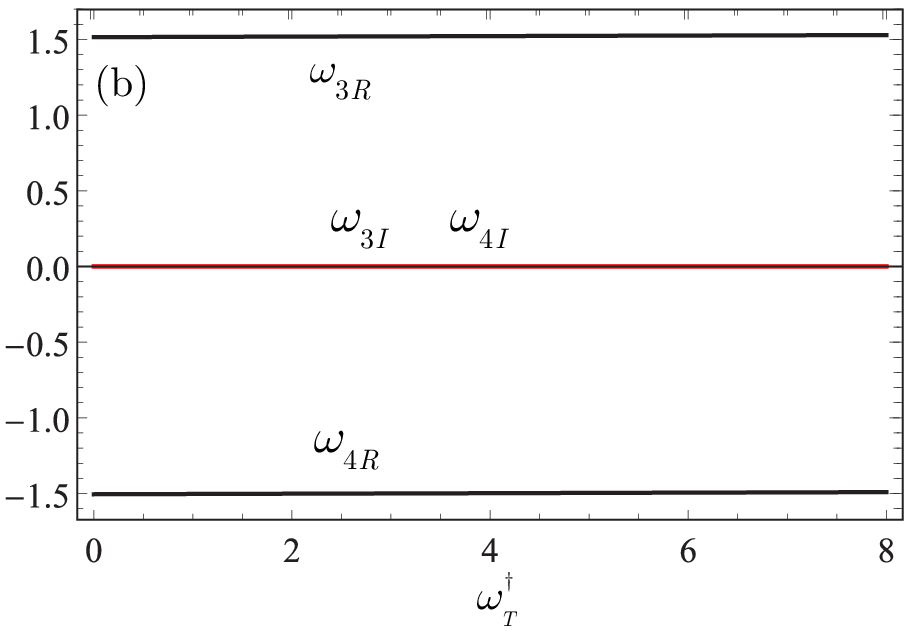}
\par\end{centering}
\caption{(a) ITG instability destabilized by spontaneous PT-symmetry breaking
through the Krein resonance when the ITG increases. (b) The electrostatic
ion cyclotron waves are stable. The gradients of temperature and density
are destabilizing factors for the electrostatic ion cyclotron waves.
However, the instability threshold imposed by PT symmetry is too high
for parameters of practical interest.}

\label{fig:ITG} 
\end{figure}

Returning to the Hamiltonian $H$ for the coupled system of drift
waves and ion cyclotron waves specified by Eq.\,(\ref{H}), we verify
that $H$ is indeed PT symmetric for 
\begin{equation}
P=\left(\begin{array}{ccccc}
1\\
 & -1\\
 &  & 1\\
 &  &  & 1\\
 &  &  &  & 1
\end{array}\right).\label{P}
\end{equation}
 This confirms that the system can only be destabilized by spontaneous
PT-symmetry breaking through the Krein resonance. The point where
the resonance occurs marks the threshold of the instability. Such
a process is displayed in Fig.\,\ref{fig:ITG}, where the real frequency
$\omega_{R}$ and the growth rate $\omega_{I}$ of the four eigenmodes
are plotted against $\omega_{T}^{\dagger}$ for a typical set of parameters
of tokamaks. The range of $\omega_{T}^{\dagger}$ is between $10^{-5}$
and $10^{-4}$. Other dimensionless parameters are $v_{thi}^{2}=4\times10^{-6}$,
$v_{s}^{2}=4\times10^{-6}$, $k_{n}=5$, $k_{y}=400$, $k_{z}=1$,
and $\gamma_{i}=1.$ It is clear that the spectrum is symmetric with
respect to the real axis as required by PT symmetry. The spontaneous
PT-symmetry breaking for the low-frequency drift waves starts at $\omega_{T}^{\dagger}=3.4\times10^{-5}$,
where two of stable drift waves resonate, and above this threshold
the drift wave is unstable. This is the familiar ITG instability.
It can be verified that the signs of actions for the two stable modes
are different at the threshold, as required by the mechanism of the
Krein resonance. For brevity, the calculation of signs of actions
for eigenmodes \citep{Zhang16GH} are omitted here. 

PT-symmetry analysis also reveals certain polarization property of
the ITG mode. When the ITG mode is stable, i.e., when PT symmetry
is unbroken, the eigenmode preserves PT symmetry,
\begin{equation}
PT\psi=P\psi^{*}=\lambda\psi\,,
\end{equation}
where $*$ denote complex conjugate. For the form of $P$ specified
in Eq.\,(\ref{P}), we can conclude that the relative phase between
$n_{i1}$ and $u_{ix1}$ needs to be locked at $\pi/2$ when the ITG
mode is stable. When the ITG instability is triggered, this relative
phase become undetermined. These characteristics might be useful for
identifying and validating the stabilization and destabilization processes
of the ITG mode in experiments \citep{Sen1991,Schmitz2016}. 

The drift wave frequency is much smaller than that of the ion cyclotron
frequency as shown in Fig.\,\ref{fig:ITG}. For drift waves, we can
neglect the $\omega^{5}$ term in Eq.\,(\ref{DR}), and the condition
for the ITG instability or spontaneous PT-symmetry breaking becomes
\begin{equation}
\beta^{2}\le4\alpha\xi\thinspace,
\end{equation}
where the equal sign holds at the threshold or the Krein resonance
point. 

In this region of parameter space, the electrostatic ion cyclotron
branch is stable without spontaneous PT-symmetry breaking. The gradients
of temperature and density are destabilizing factors for the electrostatic
ion cyclotron waves. However, for parameters of practical interest,
the threshold imposed by PT symmetry is too high for the instability
to occur. It turns out that the gradients of ion temperature and density
can drive both the electrostatic ion cyclotron modes and the drift
waves unstable without threshold, when PT-symmetry is explicitly broken
by collisions between electrons and ions, which we now investigate. 

As mentioned above, explicit PT-symmetry breaking means that the governing
equations do not admit a PT symmetry anymore due to some physical
effects, which are usually associated with dissipation. For the drift
wave dynamics studied here, one such situation arises when the collisional
effect between electrons and ions in the parallel direction is taken
into consideration. In this case, electrons' response is not purely
adiabatic, and the relationship between perturbed potential $\phi_{1}$
and $n_{e1}=Z_{i}n_{i1}$ is (in un-normalized quantities)
\begin{equation}
\phi_{1}=\left[\frac{T_{e0}}{e}+i\frac{\nu_{ie}k_{y}m_{i}\Delta u_{0}}{Z_{i}ek_{z}^{2}}\left(u_{e0}-u_{i0}\right)\right]\frac{n_{i1}}{n_{i0}}\thinspace,\label{NonAdiPhi1}
\end{equation}
where $\nu_{ie}$ is the collision frequency between ions and electrons,
and $\Delta u_{0}\equiv u_{e0}-u_{i0}$ is difference between the
equilibrium flows of electrons and ions specified by Eqs.\,(\ref{eq:ui0})
and (\ref{eq:ue0}) , respectively. The Hamiltonian of the system
is modified as
\begin{equation}
H_{\nu}=H+\left(\begin{array}{ccccc}
0 & 0 & 0 & 0 & 0\\
0 & 0 & 0 & 0 & 0\\
\frac{i\nu_{ie}\Delta u_{0}k_{y}^{2}}{k_{z}^{2}} & 0 & 0 & 0 & 0\\
0 & 0 & 0 & 0 & 0\\
0 & 0 & 0 & 0 & 0
\end{array}\right)\thinspace,\label{Hnu}
\end{equation}
where $\nu_{ie}$ has been normalized by $\Omega_{i}$ and $\Delta u_{0}$
by $a\Omega_{i}$, following the scheme in Eqs.\,(\ref{nor1})-(\ref{nor3}).
The derivation of Eqs.\,(\ref{NonAdiPhi1}) and (\ref{Hnu}) are
given in Appendix \ref{sec:Non-adiabatic-response-of}.

\begin{figure}[ptb]
\begin{centering}
\includegraphics[width=3in]{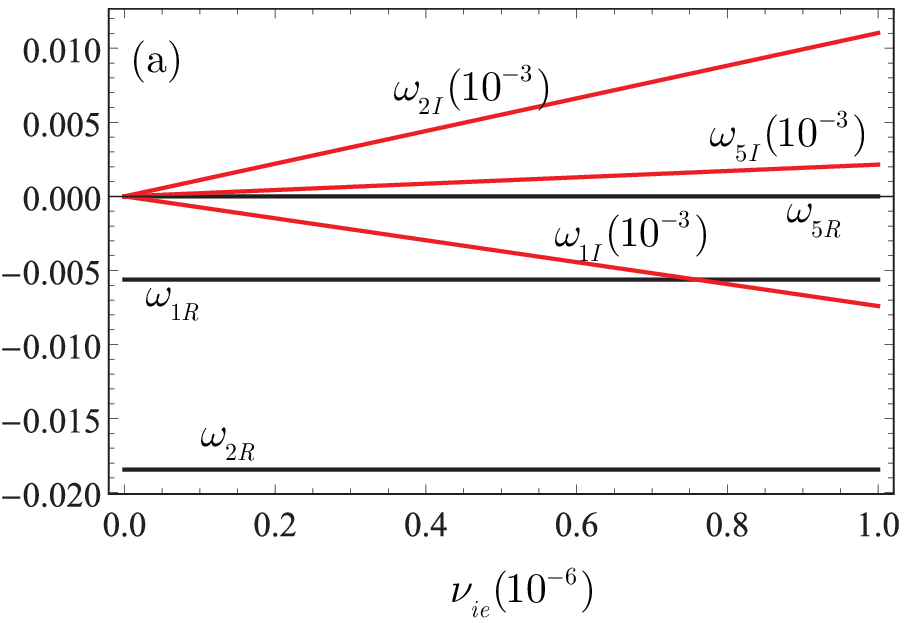}\includegraphics[width=3in]{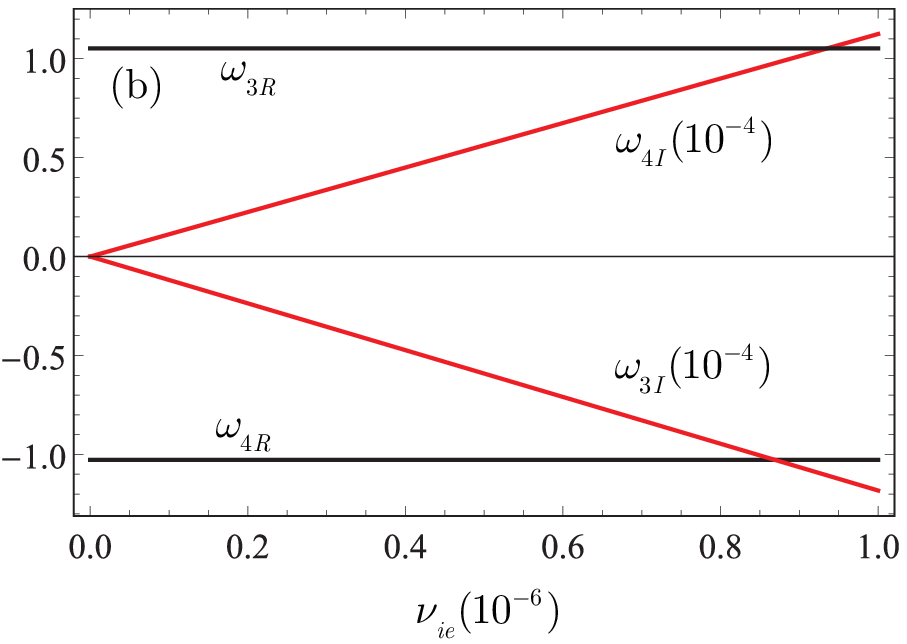}
\par\end{centering}
\caption{Drift waves (a) and electrostatic ion cyclotron waves (b) are destabilized
by the temperature and density gradients without threshold when  PT
symmetry is broken explicitly by a small but finite collisionality. }

\label{fig:collisionalMode}
\end{figure}

It can be verified that $H_{\nu}$ is not PT-symmetric, i.e., there
exists no parity operator $P$ such that $PTH=HTP$. Thus, the constraint
on the spectrum associated with PT symmetric system are removed by
the collisions between ions and electrons. An eigen frequency can
move into the complex plane without the necessity of going through
the Krein resonance first. The forbidden path for instabilities in
Fig.\,\ref{fig:Krein}(a) is allowed when PT symmetry is explicitly
broken. In this sense, the explicit PT-symmetry breaking due to dissipation
``loosens up'' the dynamics of the system, and makes the system
more susceptible to instability driving factors. In Fig.\,\ref{fig:collisionalMode},
the destabilization of the drift waves and ion cyclotron waves induced
by the finite collisionality is shown. The instabilities have no threshold.
A finite collisionality, no matter how small, will lead to a finite
growth rate. The dimensionless parameters for this case are $v_{thi}^{2}=2\times10^{-6}$,
$v_{s}^{2}=2\times10^{-6}$, $\omega_{T}^{\dagger}=10^{-4}$, $k_{n}=5$,
$k_{y}=100$, $k_{z}=1$, and $\gamma_{i}=1.$

As we see in Fig.\,\ref{fig:ITG} for the case without collisions,
ion temperature gradient can only destabilize the drift waves above
certain threshold and has no effect on the stable ion-cyclotron waves.
This can be attributed to the constraints imposed by PT-symmetry.
But when  PT symmetry is broken explicitly by a small but finite collisionality,
the temperature and density gradients destabilize both the drift waves
and the electrostatic ion cyclotron waves without threshold. 

Another noteworthy result is that finite collisionality also induces
a new low frequency drift mode which is an almost purely growing mode
with an extremely small real frequency. This mode, whose frequency
is $\omega_{5R}+i\omega_{5I}$ as in Fig.\,\ref{fig:collisionalMode}(a),
corresponds to the zero frequency mode of the dispersion relation
(\ref{DR}) when $\nu_{ie}=0$. Finite collisionality brings it to
life. Let's call it collision-induced drift wave. We are not aware
of any previous study of this mode either theoretically or experimentally. 
\begin{acknowledgments}
This research was supported by the U.S. Department of Energy (DE-AC02-09CH11466). 
\end{acknowledgments}

\bibliographystyle{apsrev4-2}
\bibliography{ITGPT}

\begin{thebibliography}{38}%
\makeatletter
\providecommand \@ifxundefined [1]{%
 \@ifx{#1\undefined}
}%
\providecommand \@ifnum [1]{%
 \ifnum #1\expandafter \@firstoftwo
 \else \expandafter \@secondoftwo
 \fi
}%
\providecommand \@ifx [1]{%
 \ifx #1\expandafter \@firstoftwo
 \else \expandafter \@secondoftwo
 \fi
}%
\providecommand \natexlab [1]{#1}%
\providecommand \enquote  [1]{``#1''}%
\providecommand \bibnamefont  [1]{#1}%
\providecommand \bibfnamefont [1]{#1}%
\providecommand \citenamefont [1]{#1}%
\providecommand \href@noop [0]{\@secondoftwo}%
\providecommand \href [0]{\begingroup \@sanitize@url \@href}%
\providecommand \@href[1]{\@@startlink{#1}\@@href}%
\providecommand \@@href[1]{\endgroup#1\@@endlink}%
\providecommand \@sanitize@url [0]{\catcode `\\12\catcode `\$12\catcode
  `\&12\catcode `\#12\catcode `\^12\catcode `\_12\catcode `\%12\relax}%
\providecommand \@@startlink[1]{}%
\providecommand \@@endlink[0]{}%
\providecommand \url  [0]{\begingroup\@sanitize@url \@url }%
\providecommand \@url [1]{\endgroup\@href {#1}{\urlprefix }}%
\providecommand \urlprefix  [0]{URL }%
\providecommand \Eprint [0]{\href }%
\providecommand \doibase [0]{https://doi.org/}%
\providecommand \selectlanguage [0]{\@gobble}%
\providecommand \bibinfo  [0]{\@secondoftwo}%
\providecommand \bibfield  [0]{\@secondoftwo}%
\providecommand \translation [1]{[#1]}%
\providecommand \BibitemOpen [0]{}%
\providecommand \bibitemStop [0]{}%
\providecommand \bibitemNoStop [0]{.\EOS\space}%
\providecommand \EOS [0]{\spacefactor3000\relax}%
\providecommand \BibitemShut  [1]{\csname bibitem#1\endcsname}%
\let\auto@bib@innerbib\@empty
\bibitem [{\citenamefont {Rudakov}\ and\ \citenamefont
  {Sagdeev}(1960)}]{Rudakov60}%
  \BibitemOpen
  \bibfield  {author} {\bibinfo {author} {\bibfnamefont {L.~I.}\ \bibnamefont
  {Rudakov}}\ and\ \bibinfo {author} {\bibfnamefont {R.}~\bibnamefont
  {Sagdeev}},\ }\href {http://www.jetp.ac.ru/cgi-bin/dn/e_010_05_0952.pdf}
  {\bibfield  {journal} {\bibinfo  {journal} {Soviet Phys. JETP}\ }\textbf
  {\bibinfo {volume} {10}},\ \bibinfo {pages} {952} (\bibinfo {year}
  {1960})}\BibitemShut {NoStop}%
\bibitem [{\citenamefont {Krall}\ and\ \citenamefont
  {Rosenbluth}(1963)}]{Krall1963}%
  \BibitemOpen
  \bibfield  {author} {\bibinfo {author} {\bibfnamefont {N.~A.}\ \bibnamefont
  {Krall}}\ and\ \bibinfo {author} {\bibfnamefont {M.~N.}\ \bibnamefont
  {Rosenbluth}},\ }\href {https://doi.org/10.1063/1.1706723} {\bibfield
  {journal} {\bibinfo  {journal} {Physics of Fluids}\ }\textbf {\bibinfo
  {volume} {6}},\ \bibinfo {pages} {254} (\bibinfo {year} {1963})}\BibitemShut
  {NoStop}%
\bibitem [{\citenamefont {Tang}(1978)}]{Tang1978}%
  \BibitemOpen
  \bibfield  {author} {\bibinfo {author} {\bibfnamefont {W.}~\bibnamefont
  {Tang}},\ }\href {https://doi.org/10.1088/0029-5515/18/8/006} {\bibfield
  {journal} {\bibinfo  {journal} {Nuclear Fusion}\ }\textbf {\bibinfo {volume}
  {18}},\ \bibinfo {pages} {1089} (\bibinfo {year} {1978})}\BibitemShut
  {NoStop}%
\bibitem [{\citenamefont {Horton}(1999)}]{horton1999drift}%
  \BibitemOpen
  \bibfield  {author} {\bibinfo {author} {\bibfnamefont {W.}~\bibnamefont
  {Horton}},\ }\href {https://doi.org/10.1103/RevModPhys.71.735} {\bibfield
  {journal} {\bibinfo  {journal} {Reviews of Modern Physics}\ }\textbf
  {\bibinfo {volume} {71}},\ \bibinfo {pages} {735} (\bibinfo {year}
  {1999})}\BibitemShut {NoStop}%
\bibitem [{\citenamefont {Horton}(2017)}]{Horton2017}%
  \BibitemOpen
  \bibfield  {author} {\bibinfo {author} {\bibfnamefont {W.}~\bibnamefont
  {Horton}},\ }\href {https://doi.org/10.1142/10595} {\emph {\bibinfo {title}
  {Turbulent Transport in Magnetized Plasmas}}}\ (\bibinfo  {publisher}
  {{WORLD} {SCIENTIFIC}},\ \bibinfo {year} {2017})\BibitemShut {NoStop}%
\bibitem [{\citenamefont {Coppi}(1967)}]{Coppi1967}%
  \BibitemOpen
  \bibfield  {author} {\bibinfo {author} {\bibfnamefont {B.}~\bibnamefont
  {Coppi}},\ }\href {https://doi.org/10.1063/1.1762151} {\bibfield  {journal}
  {\bibinfo  {journal} {Physics of Fluids}\ }\textbf {\bibinfo {volume} {10}},\
  \bibinfo {pages} {582} (\bibinfo {year} {1967})}\BibitemShut {NoStop}%
\bibitem [{\citenamefont {Parker}\ \emph {et~al.}(1993)\citenamefont {Parker},
  \citenamefont {Lee},\ and\ \citenamefont {Santoro}}]{parker1993gyrokinetic}%
  \BibitemOpen
  \bibfield  {author} {\bibinfo {author} {\bibfnamefont {S.}~\bibnamefont
  {Parker}}, \bibinfo {author} {\bibfnamefont {W.}~\bibnamefont {Lee}},\ and\
  \bibinfo {author} {\bibfnamefont {R.}~\bibnamefont {Santoro}},\ }\href
  {https://doi.org/10.1103/PhysRevLett.71.2042} {\bibfield  {journal} {\bibinfo
   {journal} {Physical Review Letters}\ }\textbf {\bibinfo {volume} {71}},\
  \bibinfo {pages} {2042} (\bibinfo {year} {1993})}\BibitemShut {NoStop}%
\bibitem [{\citenamefont {Krall}\ and\ \citenamefont
  {Rosenbluth}(1965)}]{Krall1965}%
  \BibitemOpen
  \bibfield  {author} {\bibinfo {author} {\bibfnamefont {N.~A.}\ \bibnamefont
  {Krall}}\ and\ \bibinfo {author} {\bibfnamefont {M.~N.}\ \bibnamefont
  {Rosenbluth}},\ }\href {https://doi.org/10.1063/1.1761444} {\bibfield
  {journal} {\bibinfo  {journal} {Physics of Fluids}\ }\textbf {\bibinfo
  {volume} {8}},\ \bibinfo {pages} {1488} (\bibinfo {year} {1965})}\BibitemShut
  {NoStop}%
\bibitem [{\citenamefont {Dorland}\ \emph {et~al.}(2000)\citenamefont
  {Dorland}, \citenamefont {Jenko}, \citenamefont {Kotschenreuther},\ and\
  \citenamefont {Rogers}}]{Dorland00-5579}%
  \BibitemOpen
  \bibfield  {author} {\bibinfo {author} {\bibfnamefont {W.}~\bibnamefont
  {Dorland}}, \bibinfo {author} {\bibfnamefont {F.}~\bibnamefont {Jenko}},
  \bibinfo {author} {\bibfnamefont {M.}~\bibnamefont {Kotschenreuther}},\ and\
  \bibinfo {author} {\bibfnamefont {B.~N.}\ \bibnamefont {Rogers}},\ }\href
  {https://doi.org/10.1103/PhysRevLett.85.5579} {\bibfield  {journal} {\bibinfo
   {journal} {Physical Review Letters}\ }\textbf {\bibinfo {volume} {85}},\
  \bibinfo {pages} {5579} (\bibinfo {year} {2000})}\BibitemShut {NoStop}%
\bibitem [{\citenamefont {Dimits}\ \emph {et~al.}(2007)\citenamefont {Dimits},
  \citenamefont {Nevins}, \citenamefont {Shumaker}, \citenamefont {Hammett},
  \citenamefont {Dannert}, \citenamefont {Jenko}, \citenamefont {Pueschel},
  \citenamefont {Dorland}, \citenamefont {Cowley}, \citenamefont {Leboeuf},
  \citenamefont {Rhodes}, \citenamefont {Candy},\ and\ \citenamefont
  {Estrada-Mila}}]{Dimits2007}%
  \BibitemOpen
  \bibfield  {author} {\bibinfo {author} {\bibfnamefont {A.}~\bibnamefont
  {Dimits}}, \bibinfo {author} {\bibfnamefont {W.}~\bibnamefont {Nevins}},
  \bibinfo {author} {\bibfnamefont {D.}~\bibnamefont {Shumaker}}, \bibinfo
  {author} {\bibfnamefont {G.}~\bibnamefont {Hammett}}, \bibinfo {author}
  {\bibfnamefont {T.}~\bibnamefont {Dannert}}, \bibinfo {author} {\bibfnamefont
  {F.}~\bibnamefont {Jenko}}, \bibinfo {author} {\bibfnamefont
  {M.}~\bibnamefont {Pueschel}}, \bibinfo {author} {\bibfnamefont
  {W.}~\bibnamefont {Dorland}}, \bibinfo {author} {\bibfnamefont
  {S.}~\bibnamefont {Cowley}}, \bibinfo {author} {\bibfnamefont
  {J.}~\bibnamefont {Leboeuf}}, \bibinfo {author} {\bibfnamefont
  {T.}~\bibnamefont {Rhodes}}, \bibinfo {author} {\bibfnamefont
  {J.}~\bibnamefont {Candy}},\ and\ \bibinfo {author} {\bibfnamefont
  {C.}~\bibnamefont {Estrada-Mila}},\ }\href
  {https://doi.org/10.1088/0029-5515/47/8/012} {\bibfield  {journal} {\bibinfo
  {journal} {Nuclear Fusion}\ }\textbf {\bibinfo {volume} {47}},\ \bibinfo
  {pages} {817} (\bibinfo {year} {2007})}\BibitemShut {NoStop}%
\bibitem [{\citenamefont {Hendel}(1968)}]{Hendel1968}%
  \BibitemOpen
  \bibfield  {author} {\bibinfo {author} {\bibfnamefont {H.~W.}\ \bibnamefont
  {Hendel}},\ }\href {https://doi.org/10.1063/1.1691833} {\bibfield  {journal}
  {\bibinfo  {journal} {Physics of Fluids}\ }\textbf {\bibinfo {volume} {11}},\
  \bibinfo {pages} {2426} (\bibinfo {year} {1968})}\BibitemShut {NoStop}%
\bibitem [{\citenamefont {Dirac}(1942)}]{Dirac1942}%
  \BibitemOpen
  \bibfield  {author} {\bibinfo {author} {\bibfnamefont {P.~A.}\ \bibnamefont
  {Dirac}},\ }\href {https://doi.org/10.1098/rspa.1942.0023} {\bibfield
  {journal} {\bibinfo  {journal} {Proceedings of the Royal Society of London.
  Series A. Mathematical and Physical Sciences}\ }\textbf {\bibinfo {volume}
  {180}},\ \bibinfo {pages} {1} (\bibinfo {year} {1942})}\BibitemShut {NoStop}%
\bibitem [{\citenamefont {Pauli}(1943)}]{Pauli1943}%
  \BibitemOpen
  \bibfield  {author} {\bibinfo {author} {\bibfnamefont {W.}~\bibnamefont
  {Pauli}},\ }\href {https://doi.org/10.1103/revmodphys.15.175} {\bibfield
  {journal} {\bibinfo  {journal} {Reviews of Modern Physics}\ }\textbf
  {\bibinfo {volume} {15}},\ \bibinfo {pages} {175} (\bibinfo {year}
  {1943})}\BibitemShut {NoStop}%
\bibitem [{\citenamefont {Lee}\ and\ \citenamefont {Wick}(1969)}]{Lee1969}%
  \BibitemOpen
  \bibfield  {author} {\bibinfo {author} {\bibfnamefont {T.}~\bibnamefont
  {Lee}}\ and\ \bibinfo {author} {\bibfnamefont {G.}~\bibnamefont {Wick}},\
  }\href {https://doi.org/10.1016/0550-3213(69)90098-4} {\bibfield  {journal}
  {\bibinfo  {journal} {Nuclear Physics B}\ }\textbf {\bibinfo {volume} {9}},\
  \bibinfo {pages} {209} (\bibinfo {year} {1969})}\BibitemShut {NoStop}%
\bibitem [{\citenamefont {Bender}\ and\ \citenamefont
  {Boettcher}(1998)}]{bender1998real}%
  \BibitemOpen
  \bibfield  {author} {\bibinfo {author} {\bibfnamefont {C.~M.}\ \bibnamefont
  {Bender}}\ and\ \bibinfo {author} {\bibfnamefont {S.}~\bibnamefont
  {Boettcher}},\ }\href {https://doi.org/10.1103/PhysRevLett.80.5243}
  {\bibfield  {journal} {\bibinfo  {journal} {Physical Review Letters}\
  }\textbf {\bibinfo {volume} {80}},\ \bibinfo {pages} {5243} (\bibinfo {year}
  {1998})}\BibitemShut {NoStop}%
\bibitem [{\citenamefont {Bender}\ \emph {et~al.}(2002)\citenamefont {Bender},
  \citenamefont {Brody},\ and\ \citenamefont {Jones}}]{bender2002complex}%
  \BibitemOpen
  \bibfield  {author} {\bibinfo {author} {\bibfnamefont {C.~M.}\ \bibnamefont
  {Bender}}, \bibinfo {author} {\bibfnamefont {D.~C.}\ \bibnamefont {Brody}},\
  and\ \bibinfo {author} {\bibfnamefont {H.~F.}\ \bibnamefont {Jones}},\ }\href
  {https://doi.org/10.1103/PhysRevLett.89.270401} {\bibfield  {journal}
  {\bibinfo  {journal} {Physical Review Letters}\ }\textbf {\bibinfo {volume}
  {89}},\ \bibinfo {pages} {270401} (\bibinfo {year} {2002})}\BibitemShut
  {NoStop}%
\bibitem [{\citenamefont {Bender}(2007)}]{bender2007making}%
  \BibitemOpen
  \bibfield  {author} {\bibinfo {author} {\bibfnamefont {C.~M.}\ \bibnamefont
  {Bender}},\ }\href {https://doi.org/10.1088/0034-4885/70/6/R03} {\bibfield
  {journal} {\bibinfo  {journal} {Reports on Progress in Physics}\ }\textbf
  {\bibinfo {volume} {70}},\ \bibinfo {pages} {947} (\bibinfo {year}
  {2007})}\BibitemShut {NoStop}%
\bibitem [{\citenamefont {Bender}\ and\ \citenamefont
  {Mannheim}(2010)}]{Bender2010}%
  \BibitemOpen
  \bibfield  {author} {\bibinfo {author} {\bibfnamefont {C.~M.}\ \bibnamefont
  {Bender}}\ and\ \bibinfo {author} {\bibfnamefont {P.~D.}\ \bibnamefont
  {Mannheim}},\ }\href {https://doi.org/10.1016/j.physleta.2010.02.032}
  {\bibfield  {journal} {\bibinfo  {journal} {Physics Letters A}\ }\textbf
  {\bibinfo {volume} {374}},\ \bibinfo {pages} {1616} (\bibinfo {year}
  {2010})}\BibitemShut {NoStop}%
\bibitem [{\citenamefont
  {Mostafazadeh}(2002{\natexlab{a}})}]{mostafazadeh2002pseudo}%
  \BibitemOpen
  \bibfield  {author} {\bibinfo {author} {\bibfnamefont {A.}~\bibnamefont
  {Mostafazadeh}},\ }\href {https://doi.org/10.1063/1.1418246} {\bibfield
  {journal} {\bibinfo  {journal} {Journal of Mathematical Physics}\ }\textbf
  {\bibinfo {volume} {43}},\ \bibinfo {pages} {205} (\bibinfo {year}
  {2002}{\natexlab{a}})}\BibitemShut {NoStop}%
\bibitem [{\citenamefont
  {Mostafazadeh}(2002{\natexlab{b}})}]{mostafazadeh2002pseudoII}%
  \BibitemOpen
  \bibfield  {author} {\bibinfo {author} {\bibfnamefont {A.}~\bibnamefont
  {Mostafazadeh}},\ }\href {https://doi.org/10.1063/1.1461427} {\bibfield
  {journal} {\bibinfo  {journal} {Journal of Mathematical Physics}\ }\textbf
  {\bibinfo {volume} {43}},\ \bibinfo {pages} {2814} (\bibinfo {year}
  {2002}{\natexlab{b}})}\BibitemShut {NoStop}%
\bibitem [{\citenamefont
  {Mostafazadeh}(2002{\natexlab{c}})}]{mostafazadeh2002pseudoIII}%
  \BibitemOpen
  \bibfield  {author} {\bibinfo {author} {\bibfnamefont {A.}~\bibnamefont
  {Mostafazadeh}},\ }\href {https://doi.org/10.1063/1.1489072} {\bibfield
  {journal} {\bibinfo  {journal} {Journal of Mathematical Physics}\ }\textbf
  {\bibinfo {volume} {43}},\ \bibinfo {pages} {3944} (\bibinfo {year}
  {2002}{\natexlab{c}})}\BibitemShut {NoStop}%
\bibitem [{\citenamefont {Zhang}\ \emph {et~al.}(2020)\citenamefont {Zhang},
  \citenamefont {Qin},\ and\ \citenamefont {Xiao}}]{Zhang2020PT}%
  \BibitemOpen
  \bibfield  {author} {\bibinfo {author} {\bibfnamefont {R.}~\bibnamefont
  {Zhang}}, \bibinfo {author} {\bibfnamefont {H.}~\bibnamefont {Qin}},\ and\
  \bibinfo {author} {\bibfnamefont {J.}~\bibnamefont {Xiao}},\ }\href
  {https://doi.org/10.1063/1.5117211} {\bibfield  {journal} {\bibinfo
  {journal} {Journal of Mathematical Physics}\ }\textbf {\bibinfo {volume}
  {61}},\ \bibinfo {pages} {012101} (\bibinfo {year} {2020})}\BibitemShut
  {NoStop}%
\bibitem [{\citenamefont {Qin}\ \emph {et~al.}(2019)\citenamefont {Qin},
  \citenamefont {Zhang}, \citenamefont {Glasser},\ and\ \citenamefont
  {Xiao}}]{Qin2019KH}%
  \BibitemOpen
  \bibfield  {author} {\bibinfo {author} {\bibfnamefont {H.}~\bibnamefont
  {Qin}}, \bibinfo {author} {\bibfnamefont {R.}~\bibnamefont {Zhang}}, \bibinfo
  {author} {\bibfnamefont {A.~S.}\ \bibnamefont {Glasser}},\ and\ \bibinfo
  {author} {\bibfnamefont {J.}~\bibnamefont {Xiao}},\ }\href
  {https://doi.org/10.1063/1.5088498} {\bibfield  {journal} {\bibinfo
  {journal} {Physics of Plasmas}\ }\textbf {\bibinfo {volume} {26}},\ \bibinfo
  {pages} {032102} (\bibinfo {year} {2019})}\BibitemShut {NoStop}%
\bibitem [{\citenamefont {Fu}\ and\ \citenamefont {Qin}(2020)}]{Fu2020KH}%
  \BibitemOpen
  \bibfield  {author} {\bibinfo {author} {\bibfnamefont {Y.}~\bibnamefont
  {Fu}}\ and\ \bibinfo {author} {\bibfnamefont {H.}~\bibnamefont {Qin}},\
  }\href {https://doi.org/10.1088/1367-2630/aba38f} {\bibfield  {journal}
  {\bibinfo  {journal} {New Journal of Physics}\ }\textbf {\bibinfo {volume}
  {22}},\ \bibinfo {pages} {083040} (\bibinfo {year} {2020})}\BibitemShut
  {NoStop}%
\bibitem [{\citenamefont {Krein}(1950)}]{Krein50}%
  \BibitemOpen
  \bibfield  {author} {\bibinfo {author} {\bibfnamefont {M.}~\bibnamefont
  {Krein}},\ }\href@noop {} {\bibfield  {journal} {\bibinfo  {journal} {Doklady
  Akad. Nauk. SSSR N.S.}\ }\textbf {\bibinfo {volume} {73}},\ \bibinfo {pages}
  {445} (\bibinfo {year} {1950})}\BibitemShut {NoStop}%
\bibitem [{\citenamefont {Gel'fand}\ and\ \citenamefont
  {Lidskii}(1955)}]{Gelfand55}%
  \BibitemOpen
  \bibfield  {author} {\bibinfo {author} {\bibfnamefont {I.~M.}\ \bibnamefont
  {Gel'fand}}\ and\ \bibinfo {author} {\bibfnamefont {V.~B.}\ \bibnamefont
  {Lidskii}},\ }\href@noop {} {\bibfield  {journal} {\bibinfo  {journal}
  {Uspekhi Mat. Nauk}\ }\textbf {\bibinfo {volume} {10}},\ \bibinfo {pages} {3}
  (\bibinfo {year} {1955})}\BibitemShut {NoStop}%
\bibitem [{\citenamefont {Yakubovich}\ and\ \citenamefont
  {Starzhinskii}(1975)}]{Yakubovich75}%
  \BibitemOpen
  \bibfield  {author} {\bibinfo {author} {\bibfnamefont {V.}~\bibnamefont
  {Yakubovich}}\ and\ \bibinfo {author} {\bibfnamefont {V.}~\bibnamefont
  {Starzhinskii}},\ }\href@noop {} {\emph {\bibinfo {title} {Linear
  Differential Equations with Periodic Coefficients}}},\ Vol.~\bibinfo {volume}
  {I}\ (\bibinfo  {publisher} {Wiley},\ \bibinfo {address} {New York},\
  \bibinfo {year} {1975})\BibitemShut {NoStop}%
\bibitem [{\citenamefont {Kirillov}(2013)}]{Kirillov2013a}%
  \BibitemOpen
  \bibfield  {author} {\bibinfo {author} {\bibfnamefont {O.~N.}\ \bibnamefont
  {Kirillov}},\ }\href {https://doi.org/10.1515/9783110270433} {\emph {\bibinfo
  {title} {Nonconservative stability problems of modern physics}}},\ \bibinfo
  {series} {Studies in Mathematical Physics}, Vol.~\bibinfo {volume} {14}\
  (\bibinfo  {publisher} {Walter de Gruyter},\ \bibinfo {year}
  {2013})\BibitemShut {NoStop}%
\bibitem [{\citenamefont {Parker}\ \emph {et~al.}(2020)\citenamefont {Parker},
  \citenamefont {Marston}, \citenamefont {Tobias},\ and\ \citenamefont
  {Zhu}}]{Parker2020}%
  \BibitemOpen
  \bibfield  {author} {\bibinfo {author} {\bibfnamefont {J.~B.}\ \bibnamefont
  {Parker}}, \bibinfo {author} {\bibfnamefont {J.}~\bibnamefont {Marston}},
  \bibinfo {author} {\bibfnamefont {S.~M.}\ \bibnamefont {Tobias}},\ and\
  \bibinfo {author} {\bibfnamefont {Z.}~\bibnamefont {Zhu}},\ }\href
  {https://doi.org/10.1103/physrevlett.124.195001} {\bibfield  {journal}
  {\bibinfo  {journal} {Physical Review Letters}\ }\textbf {\bibinfo {volume}
  {124}},\ \bibinfo {pages} {195001} (\bibinfo {year} {2020})}\BibitemShut
  {NoStop}%
\bibitem [{\citenamefont {Bender}(2019)}]{BenderPrivate2019}%
  \BibitemOpen
  \bibfield  {author} {\bibinfo {author} {\bibfnamefont {C.~M.}\ \bibnamefont
  {Bender}},\ }\href@noop {} {} (\bibinfo {year} {2019}),\ \bibinfo {note}
  {private communication}\BibitemShut {NoStop}%
\bibitem [{\citenamefont {Zhang}\ \emph {et~al.}(2016)\citenamefont {Zhang},
  \citenamefont {Qin}, \citenamefont {Davidson}, \citenamefont {Liu},\ and\
  \citenamefont {Xiao}}]{Zhang16GH}%
  \BibitemOpen
  \bibfield  {author} {\bibinfo {author} {\bibfnamefont {R.}~\bibnamefont
  {Zhang}}, \bibinfo {author} {\bibfnamefont {H.}~\bibnamefont {Qin}}, \bibinfo
  {author} {\bibfnamefont {R.~C.}\ \bibnamefont {Davidson}}, \bibinfo {author}
  {\bibfnamefont {J.}~\bibnamefont {Liu}},\ and\ \bibinfo {author}
  {\bibfnamefont {J.}~\bibnamefont {Xiao}},\ }\href
  {https://doi.org/10.1063/1.4954832} {\bibfield  {journal} {\bibinfo
  {journal} {Physics of Plasmas}\ }\textbf {\bibinfo {volume} {23}},\ \bibinfo
  {pages} {072111} (\bibinfo {year} {2016})}\BibitemShut {NoStop}%
\bibitem [{\citenamefont {Qin}(2019)}]{Qin2019LH}%
  \BibitemOpen
  \bibfield  {author} {\bibinfo {author} {\bibfnamefont {H.}~\bibnamefont
  {Qin}},\ }\href {https://doi.org/10.1063/1.5067391} {\bibfield  {journal}
  {\bibinfo  {journal} {Journal of Mathematical Physics}\ }\textbf {\bibinfo
  {volume} {60}},\ \bibinfo {pages} {022901} (\bibinfo {year}
  {2019})}\BibitemShut {NoStop}%
\bibitem [{\citenamefont {Qin}\ \emph {et~al.}(2014)\citenamefont {Qin},
  \citenamefont {Davidson}, \citenamefont {Burby},\ and\ \citenamefont
  {Chung}}]{Qin14-044001}%
  \BibitemOpen
  \bibfield  {author} {\bibinfo {author} {\bibfnamefont {H.}~\bibnamefont
  {Qin}}, \bibinfo {author} {\bibfnamefont {R.~C.}\ \bibnamefont {Davidson}},
  \bibinfo {author} {\bibfnamefont {J.~W.}\ \bibnamefont {Burby}},\ and\
  \bibinfo {author} {\bibfnamefont {M.}~\bibnamefont {Chung}},\ }\href
  {https://doi.org/10.1103/PhysRevSTAB.17.044001} {\bibfield  {journal}
  {\bibinfo  {journal} {Physical Review Special Topics - Accelerators and
  Beams}\ }\textbf {\bibinfo {volume} {17}},\ \bibinfo {pages} {044001}
  (\bibinfo {year} {2014})}\BibitemShut {NoStop}%
\bibitem [{\citenamefont {Qin}\ \emph {et~al.}(2015)\citenamefont {Qin},
  \citenamefont {Chung}, \citenamefont {Davidson},\ and\ \citenamefont
  {Burby}}]{Qin15-056702}%
  \BibitemOpen
  \bibfield  {author} {\bibinfo {author} {\bibfnamefont {H.}~\bibnamefont
  {Qin}}, \bibinfo {author} {\bibfnamefont {M.}~\bibnamefont {Chung}}, \bibinfo
  {author} {\bibfnamefont {R.~C.}\ \bibnamefont {Davidson}},\ and\ \bibinfo
  {author} {\bibfnamefont {J.~W.}\ \bibnamefont {Burby}},\ }\href
  {https://doi.org/10.1063/1.4920961} {\bibfield  {journal} {\bibinfo
  {journal} {Physics of Plasmas}\ }\textbf {\bibinfo {volume} {22}},\ \bibinfo
  {pages} {056702} (\bibinfo {year} {2015})}\BibitemShut {NoStop}%
\bibitem [{\citenamefont {Chung}\ \emph {et~al.}(2015)\citenamefont {Chung},
  \citenamefont {Qin}, \citenamefont {Groening}, \citenamefont {Davidson},\
  and\ \citenamefont {Xiao}}]{Chung15}%
  \BibitemOpen
  \bibfield  {author} {\bibinfo {author} {\bibfnamefont {M.}~\bibnamefont
  {Chung}}, \bibinfo {author} {\bibfnamefont {H.}~\bibnamefont {Qin}}, \bibinfo
  {author} {\bibfnamefont {L.}~\bibnamefont {Groening}}, \bibinfo {author}
  {\bibfnamefont {R.~C.}\ \bibnamefont {Davidson}},\ and\ \bibinfo {author}
  {\bibfnamefont {C.}~\bibnamefont {Xiao}},\ }\href
  {https://doi.org/10.1063/1.4903457} {\bibfield  {journal} {\bibinfo
  {journal} {Physics of Plasmas}\ }\textbf {\bibinfo {volume} {22}},\ \bibinfo
  {pages} {013109} (\bibinfo {year} {2015})}\BibitemShut {NoStop}%
\bibitem [{\citenamefont {Fraser}\ \emph {et~al.}(2018)\citenamefont {Fraser},
  \citenamefont {Pueschel}, \citenamefont {Terry},\ and\ \citenamefont
  {Zweibel}}]{Fraser2018}%
  \BibitemOpen
  \bibfield  {author} {\bibinfo {author} {\bibfnamefont {A.~E.}\ \bibnamefont
  {Fraser}}, \bibinfo {author} {\bibfnamefont {M.~J.}\ \bibnamefont
  {Pueschel}}, \bibinfo {author} {\bibfnamefont {P.~W.}\ \bibnamefont
  {Terry}},\ and\ \bibinfo {author} {\bibfnamefont {E.~G.}\ \bibnamefont
  {Zweibel}},\ }\href {https://doi.org/10.1063/1.5049580} {\bibfield  {journal}
  {\bibinfo  {journal} {Physics of Plasmas}\ }\textbf {\bibinfo {volume}
  {25}},\ \bibinfo {pages} {122303} (\bibinfo {year} {2018})}\BibitemShut
  {NoStop}%
\bibitem [{\citenamefont {Sen}\ \emph {et~al.}(1991)\citenamefont {Sen},
  \citenamefont {Chen},\ and\ \citenamefont {Mauel}}]{Sen1991}%
  \BibitemOpen
  \bibfield  {author} {\bibinfo {author} {\bibfnamefont {A.}~\bibnamefont
  {Sen}}, \bibinfo {author} {\bibfnamefont {J.}~\bibnamefont {Chen}},\ and\
  \bibinfo {author} {\bibfnamefont {M.}~\bibnamefont {Mauel}},\ }\href
  {https://doi.org/10.1103/physrevlett.66.429} {\bibfield  {journal} {\bibinfo
  {journal} {Physical Review Letters}\ }\textbf {\bibinfo {volume} {66}},\
  \bibinfo {pages} {429} (\bibinfo {year} {1991})}\BibitemShut {NoStop}%
\bibitem [{\citenamefont {Schmitz}\ \emph {et~al.}(2016)\citenamefont
  {Schmitz}, \citenamefont {Fulton}, \citenamefont {Ruskov}, \citenamefont
  {Lau}, \citenamefont {Deng}, \citenamefont {Tajima}, \citenamefont
  {Binderbauer}, \citenamefont {Holod}, \citenamefont {Lin}, \citenamefont
  {Gota}, \citenamefont {Tuszewski}, \citenamefont {Dettrick},\ and\
  \citenamefont {Steinhauer}}]{Schmitz2016}%
  \BibitemOpen
  \bibfield  {author} {\bibinfo {author} {\bibfnamefont {L.}~\bibnamefont
  {Schmitz}}, \bibinfo {author} {\bibfnamefont {D.~P.}\ \bibnamefont {Fulton}},
  \bibinfo {author} {\bibfnamefont {E.}~\bibnamefont {Ruskov}}, \bibinfo
  {author} {\bibfnamefont {C.}~\bibnamefont {Lau}}, \bibinfo {author}
  {\bibfnamefont {B.~H.}\ \bibnamefont {Deng}}, \bibinfo {author}
  {\bibfnamefont {T.}~\bibnamefont {Tajima}}, \bibinfo {author} {\bibfnamefont
  {M.~W.}\ \bibnamefont {Binderbauer}}, \bibinfo {author} {\bibfnamefont
  {I.}~\bibnamefont {Holod}}, \bibinfo {author} {\bibfnamefont
  {Z.}~\bibnamefont {Lin}}, \bibinfo {author} {\bibfnamefont {H.}~\bibnamefont
  {Gota}}, \bibinfo {author} {\bibfnamefont {M.}~\bibnamefont {Tuszewski}},
  \bibinfo {author} {\bibfnamefont {S.~A.}\ \bibnamefont {Dettrick}},\ and\
  \bibinfo {author} {\bibfnamefont {L.~C.}\ \bibnamefont {Steinhauer}},\ }\href
  {https://doi.org/10.1038/ncomms13860} {\bibfield  {journal} {\bibinfo
  {journal} {Nature Communications}\ }\textbf {\bibinfo {volume} {7}},\
  \bibinfo {pages} {13860} (\bibinfo {year} {2016})}\BibitemShut {NoStop}%
\end{thebibliography}%

\appendix

\section{Non-adiabatic response of electrons due to collisions\label{sec:Non-adiabatic-response-of}}

In this appendix, we derive electrons' non-adiabatic response to the
perturbed potential when the collisions between electrons and ions
are taken into account within the two-fluid model. The derivation
is given using un-normalized quantities. 

From the fluid equations, the adiabatic electron response can be obtained
from the perturbed parallel momentum equation of electrons,
\begin{equation}
eik_{z}\phi_{1}-ik_{z}T_{e0}\frac{n_{e1}}{n_{e0}}=0\thinspace,\label{parallmomentum}
\end{equation}
where the electron inertial and temperature perturbation have been
neglected. To model the non-adiabatic electron response due to electron-ion
collisions, we augment Eq.\,(\ref{parallmomentum}) with a term describing
the momentum exchange induced by the collisions,
\begin{equation}
eik_{z}\phi_{1}-ik_{z}T_{e0}\frac{n_{e1}}{n_{e0}}-m_{e}\nu_{ei}\left(u_{e1z}-u_{i1z}\right)=0\thinspace.\label{kzphi1}
\end{equation}
Similarly, the perturbed parallel momentum equation for ions (\ref{uiz1})
is modified to
\begin{equation}
-i\omega u_{iz1}=-\frac{ik_{z}Z_{i}e\phi_{1}}{m_{i}}-\frac{ik_{z}p_{i1}}{m_{i}n_{i0}}-\nu_{ie}\left(u_{iz1}-u_{ez1}\right)\thinspace.\label{uiz1-mod}
\end{equation}
Momentum conservation requires that $\nu_{ie}=\nu_{ei}Z_{i}m_{e}/m_{i}$.
To obtain a closed expression for the collision term $m_{e}\nu_{ei}\left(u_{e1z}-u_{i1z}\right)$,
we look at the continuity equation for both species,
\begin{align}
-i\omega n_{i1} & =-ik_{y}u_{i0}n_{i1}-ik_{z}n_{i0}u_{i1z}+\frac{ick_{y}}{B}\frac{dn_{i0}}{dx}\phi_{1}\thinspace,\label{ni1-approx}\\
-i\omega n_{e1} & =-ik_{y}u_{e0}n_{e1}-ik_{z}n_{e0}u_{e1z}+\frac{ick_{y}}{B}\frac{dn_{e0}}{dx}\phi_{1}\,,\label{ne1-approx}
\end{align}
where the perturbed perpendicular flows for both species have been
approximated by the $\boldsymbol{E}\times\boldsymbol{B}$ flow due
to $\phi_{1}$. With the quasi-neutrality condition, Eqs.\,(\ref{ni1-approx})
and (\ref{ne1-approx}) lead to 
\begin{equation}
u_{e1z}-u_{i1z}=-\frac{k_{y}}{k_{z}}\left(u_{e0}-u_{i0}\right)\frac{n_{i1}}{n_{i0}}\thinspace.\label{du1}
\end{equation}
Substituting Eq.\,(\ref{du1}) into Eq.\,(\ref{kzphi1}), we obtain
the non-adiabatic response of electrons,
\begin{equation}
\phi_{1}=\left[\frac{T_{e0}}{e}+i\frac{\nu_{ei}k_{y}m_{e}\Delta u_{0}}{ek_{z}^{2}}\left(u_{e0}-u_{i0}\right)\right]\frac{n_{i1}}{n_{i0}}\thinspace,\label{NonAdiPhi1-2}
\end{equation}
which is Eq.\,(\ref{NonAdiPhi1}). This non-adiabatic response only
modifies one element of the Hamiltonian matrix $H$, i.e., the (3,1)
element. For Eq.\,(\ref{uiz1-mod}), the non-adiabatic part of $\phi_{1}$,
i.e, the second term on the right-hand side of Eq.\,(\ref{NonAdiPhi1-2})
cancels with the collision term, and the perturbed parallel momentum
equation for ions remains the same as the collisionless case. After
normalization using the scheme listed in Eqs.\,(\ref{nor1})-(\ref{nor3}),
we obtain the Hamiltonian matrix $H_{\nu}$ given by Eq.\,(\ref{Hnu}). 
\end{document}